\documentclass[reprint,amsmath,amssymb,aps,floatfix,superscriptaddress,longbibliography]{revtex4-2}
\usepackage{graphicx}
\usepackage{hyperref}

\newcommand*\mean[1]{\langle #1 \rangle}
\newcommand*\tr{\tilde{r}}
\newcommand*\tn{\tilde{n}}

\begin{document}
\title{Observation of scale invariance in two-dimensional matter-wave Townes solitons}
\author{Cheng-An Chen}
\affiliation{Department of Physics and Astronomy, Purdue University, West Lafayette, IN 47907, USA}
\author{Chen-Lung Hung}
\email{clhung@purdue.edu}
\affiliation{Department of Physics and Astronomy, Purdue University, West Lafayette, IN 47907, USA}
\affiliation{Purdue Quantum Science and Engineering Institute, Purdue University, West Lafayette, IN 47907, USA}
\date{\today}

\begin{abstract}
We report near-deterministic generation of two-dimensional (2D) matter-wave Townes solitons, and a precision test on scale invariance in attractive 2D Boses gases. We induce a shape-controlled modulational instability in an elongated 2D matter-wave to create an array of isolated solitary waves of various sizes and peak densities. We confirm scale invariance by observing the collapse of solitary-wave density profiles onto a single curve in a dimensionless coordinate rescaled according to their peak densities, and observe that the scale-invariant profiles measured at different coupling constants $g$ can further collapse onto the universal profile of Townes solitons. The reported scaling behavior is tested with a nearly 60-fold difference in soliton interaction energies, and allows us to discuss the impact of a non-negligible magnetic dipole-dipole interaction (MDDI) on 2D scale invariance. We confirm that the effect of MDDI in our alkali cesium quasi-2D samples effectively conforms to the same scaling law governed by a contact interaction to well within our experiment uncertainty.
\end{abstract}
\maketitle

A scale-invariant system possesses self-similar features that can occur at all scales, where system observables exhibit general scaling behaviors. Weakly interacting two-dimensional (2D) Bose gases offer unique opportunities to explore scale invariance (SI) in a many-body system, because the effective contact interaction potential and single-particle dispersion both have the same scale dependence \cite{pitaevskii1997breathing,posazhennikova2006colloquium}. The ability to tune the contact interaction strength $g$ via a magnetic Feshbach resonance \cite{chin2010feshbach} further allows for explorations of SI over a wide parameter range, both in equilibrium and from out-of-equilibrium dynamics. At repulsive interactions ($g>0$), SI has been observed in density observables associated with the equations of states, in normal and superfluid phases, and across the Berezinskii-Kosterlitz-Thouless superfluid phase transition, offering a rich understanding of scale-invariant 2D many-body phases \cite{hung2011observation,yefsah2011exploring,desbuquois2012superfluid,ranccon2012universal,ha2013strongly,desbuquois2014determination}. However, 2D Bose gases with attractive interactions ($g<0$) have rarely been studied primarily due to an instability to collapse under typical experiment trap conditions \cite{kagan1998collapse,donley2001dynamics}. When and how does SI manifest in the unstable attractive regime has remained relatively unexplored.

One intriguing example occurs deep in quantum degeneracy, when attractive 2D Bose gases form matter-waves that may sustain a scale-invariant, quasi-stationary state -- a prediction originally made for self-focusing optical beams called the Townes soliton \cite{chiao1964self}. Under SI, a Townes soliton may form at any length scale $\lambda$, but only under an isotropic wave function $\psi (\mathbf{r}) = \phi(r/\lambda)/\lambda$, where $\phi(\tr)$ is a dimensionless Townes profile \cite{SM}. The atom number in a Townes soliton is necessarily fixed at $N_\mathrm{ts} =\int |\phi(\tr)|^2d\tilde{\mathbf{r}}\approx 5.85/|g|$. At this atom number the matter-wave dispersion intricately balances against the mean field attraction. The main challenge for realizing scale-invariant 2D solitons is that they are unstable \cite{pedri2005two,kartashov2011solitons}, and have not been realized in equilibrium. In nonlinear optics, a Townes profile has been partially observed in a collapsed optical wave \cite{moll2003self}.

\begin{figure}[b]
\centering
\includegraphics[width=1\columnwidth]{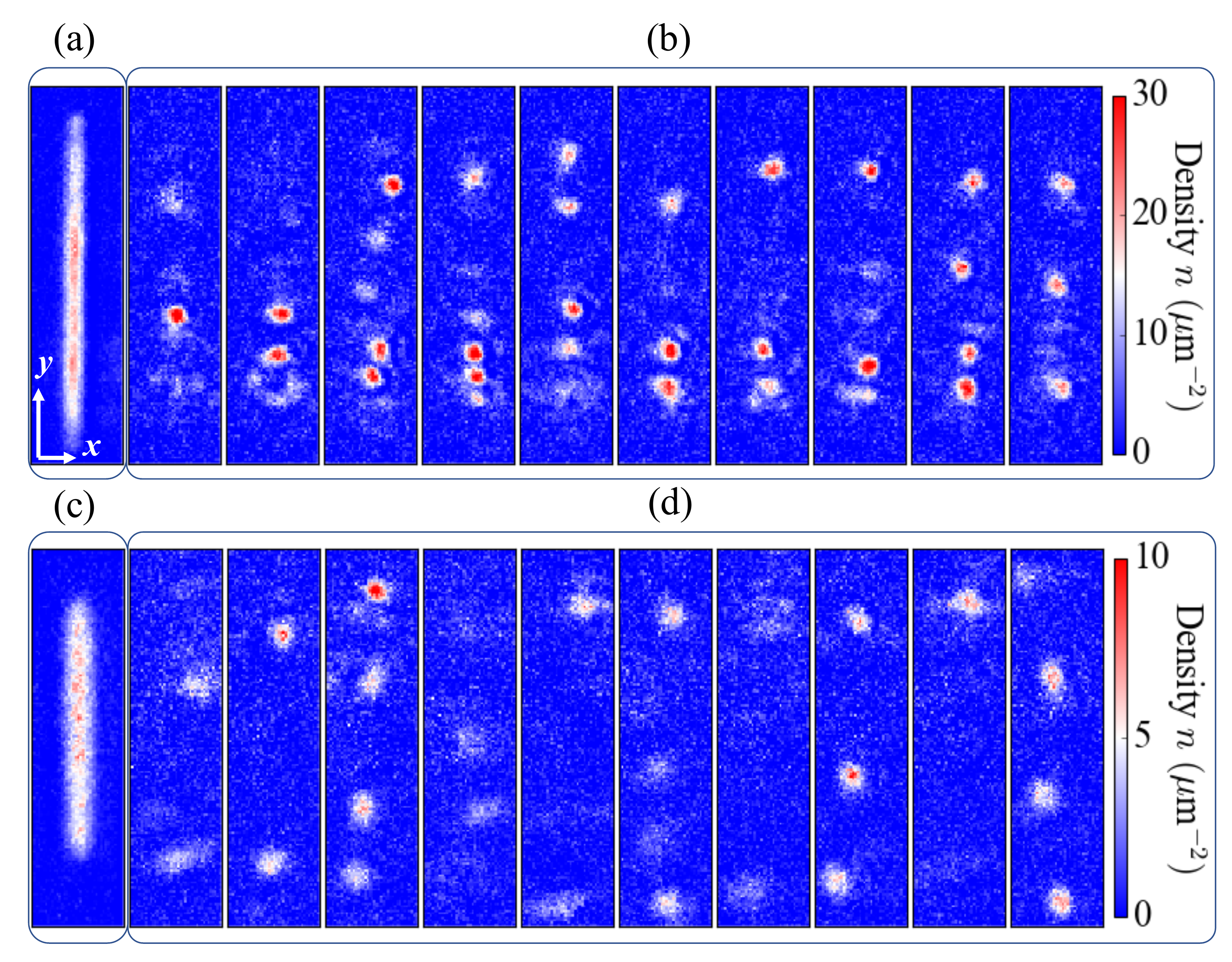}
\caption{Formation of 2D matter-wave soliton trains. (a) An elongated 2D Bose gas of peak density $n_i\approx 20/\mu$m$^{2}$ is held at an initial coupling constant $g_i\approx 0.129$ and quenched to a new coupling constant $g\approx -0.0215$, with simultaneous removal of the horizontal confinement in the $x$-$y$ plane. Arrays of solitary waves are observed in shot-to-shot images in (b), taken after a $50$~ms wait time. A different sample in (c) is prepared at a much lower initial peak density $n_i\approx6/\mu$m$^2$ and quenched to $g\approx -0.0075$, similarly generating solitary waves as observed in (d). Image size in (a,b): 19$\times$77 $\mu$m$^2$. Image size in (c,d): 40$\times$160 $\mu$m$^2$.}
\label{fig:demo}
\end{figure}

To date, an experimental demonstration of SI in 2D matter-wave solitons has remained elusive. Recently in Ref.~\cite{chen2020observation}, it is observed that an interaction quench in a homogeneous 2D superfluid to $g<0$ can induce a modulational instability (MI) \cite{nguyen2017formation}, which fragments a large sample into many density blobs with atom numbers universally around $N_\mathrm{ts}$. Townes solitons of similar peak densities (and sizes) are observed to form randomly from the blobs. However, dispersion, collisions and collapse of many blobs generate remnants throughout a large sample, making confirmation of SI in solitons a nontrivial task. Besides soliton formation in quench dynamics, an optical technique \cite{zou2021optical} has been developed very recently to deterministically imprint a Townes soliton in a two-component planar Bose gas \cite{bakkalihassani2021realization}.

In this letter, we report a simple recipe to create isolated 2D solitons with peak densities differing by 20-fold, thus enabling unambiguous experimental tests on SI. Our method induces controlled MI in an elongated 2D superfluid that fragments into an array of solitary waves nearly free from background remnants. Using these samples, we confirm SI by observing their density profiles collapse onto a single curve in a dimensionless coordinate $\tilde{r}=\sqrt{n_p}r$, where $n_p$ is the peak density that sets the length scale $\lambda=1/\sqrt{n_p}$. We further confirm that the scale-invariant density profiles measured at different coupling constants $g$ can collapse onto a universal curve, which agrees remarkably well with the Townes profile. Furthermore, we discuss the effect of a non-local MDDI in our quasi-2D geometry, which conforms to the same scaling law governed by a contact interaction to well within our experiment uncertainty.

Our experiment begins with a 2D superfluid formed by a variable number of cesium atoms ($N\approx 6\times 10^3\sim 1.5\times 10^4$) polarized in the $|F=3, m_F=3\rangle$ hyperfine ground state and with a low temperature $T\lesssim 8~$nK. The superfluid is trapped inside a quasi-2D box potential formed by all repulsive optical dipole beams with an adjustable horizontal box confinement. The tight vertical ($z$-) confinement freezes all atoms in the harmonic ground state along the imaging axis, giving a trap vibrational frequency $\omega_z = 2\pi \times 2.25(1)~$kHz and a harmonic oscillator length $l_z \approx 184~$nm. The 2D coupling constant $g =\sqrt{8\pi}a/l_z$ is controlled by a tunable s-wave scattering length $a$, initially prepared at $g=g_i \approx 0.129$ and later quenched to a negative value $g<0$ via a magnetic Feshbach resonance \cite{chin2010feshbach}. The coupling constant is calibrated with an uncertainty $\delta g \approx \pm 0.0005$ \cite{SM}. Following the interaction quench and simultaneous removal of the horizontal box confinement, the 2D gas is allowed to evolve freely in the horizontal plane for a hold time of $\sim 50~$ms, which is sufficiently long to allow samples to fully fragment but short enough so that there is not a significant atom loss that could make a soliton unstable. Absorption imaging is then performed to record the density distribution; see Fig.~\ref{fig:demo} for sample images. The image resolution is experimentally determined to be $\sim 1.5~\mu$m ($1/e^2$ Gaussian width) \cite{hung2011extracting,chen2021observation}.

\begin{figure}[t]
\centering
\includegraphics[width=1\columnwidth]{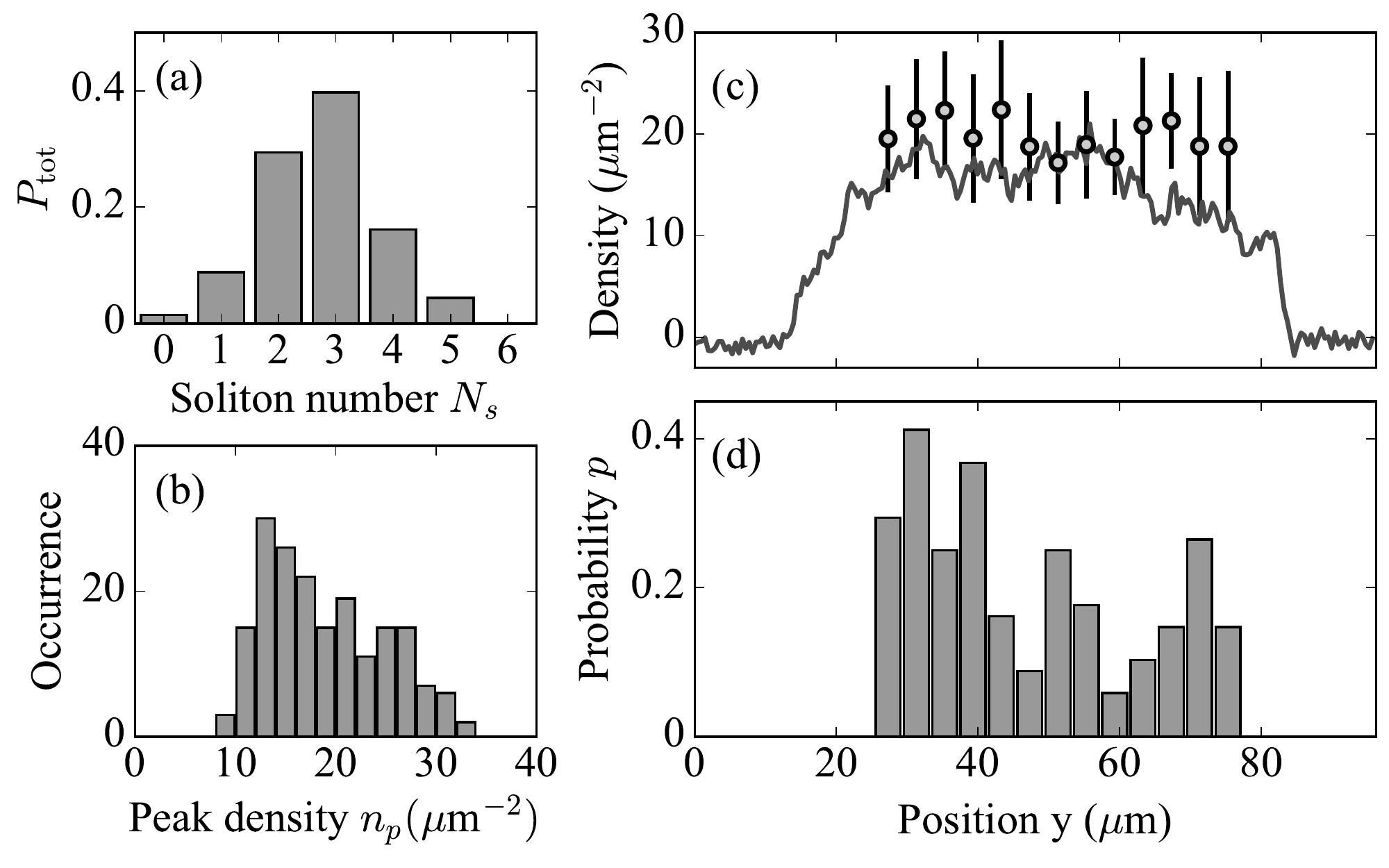}
\caption{Soliton formation statistics. (a) Probability $P_\mathrm{tot}$ of finding $N_\mathrm{s}$ solitons after the quench, evaluated using 68 samples as shown in Fig.~\ref{fig:demo}(b). (b) Occurrence of solitons with peak density $n_p$ (Bin size: $2/\mu$m$^2$). (c) Average peak density $\bar{n}_p$ versus position along the long ($y$-)axis (filled circles). Error bars represent standard deviation. Solid curve shows the density $n_i$ of the initial sample through the long axis. (d) Probability for observing a soliton at position $y$ in a quenched sample (Bin size: $4~\mu$m).}
\label{fig:statistics}
\end{figure}

\begin{figure}[t]
\centering
\includegraphics[width=1\columnwidth]{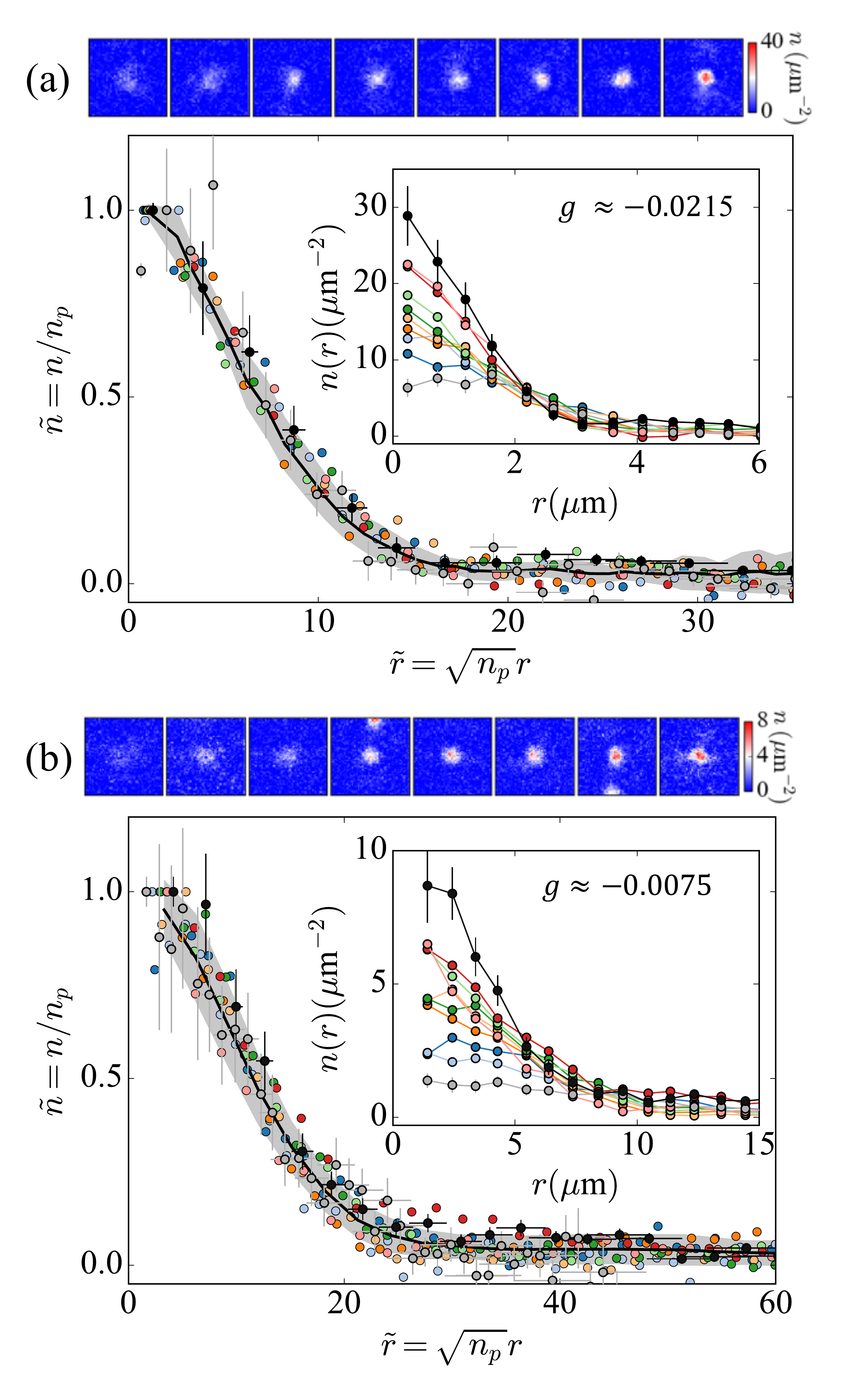}
\caption{Testing scale invariance. (a) Images in the top panel, from left to right, show solitons of low to high peak densities, selected from samples as shown in Fig.~\ref{fig:demo}(b). Image size: $19 \times 19\mu$m$^2$. Their radial density profiles $n(r)$ (filled circles, inset) approximately collapse onto a single curve in the rescaled coordinate $\tilde{r} = \sqrt{n_p}r$ and $\tilde{n} = n/n_p$. Error bars include statistical and systematic errors. Shaded band shows the standard deviation of 20 rescaled radial profiles around their mean $\mean{\tilde{n}}$ (solid curve). (b) similarly shows soliton images and profiles observed in Fig.~\ref{fig:demo}(d). Image size: $60 \times 60\mu$m$^2$.}
\label{fig:invariant}
\end{figure}

\begin{figure}[t]
\centering
\includegraphics[width=1\columnwidth]{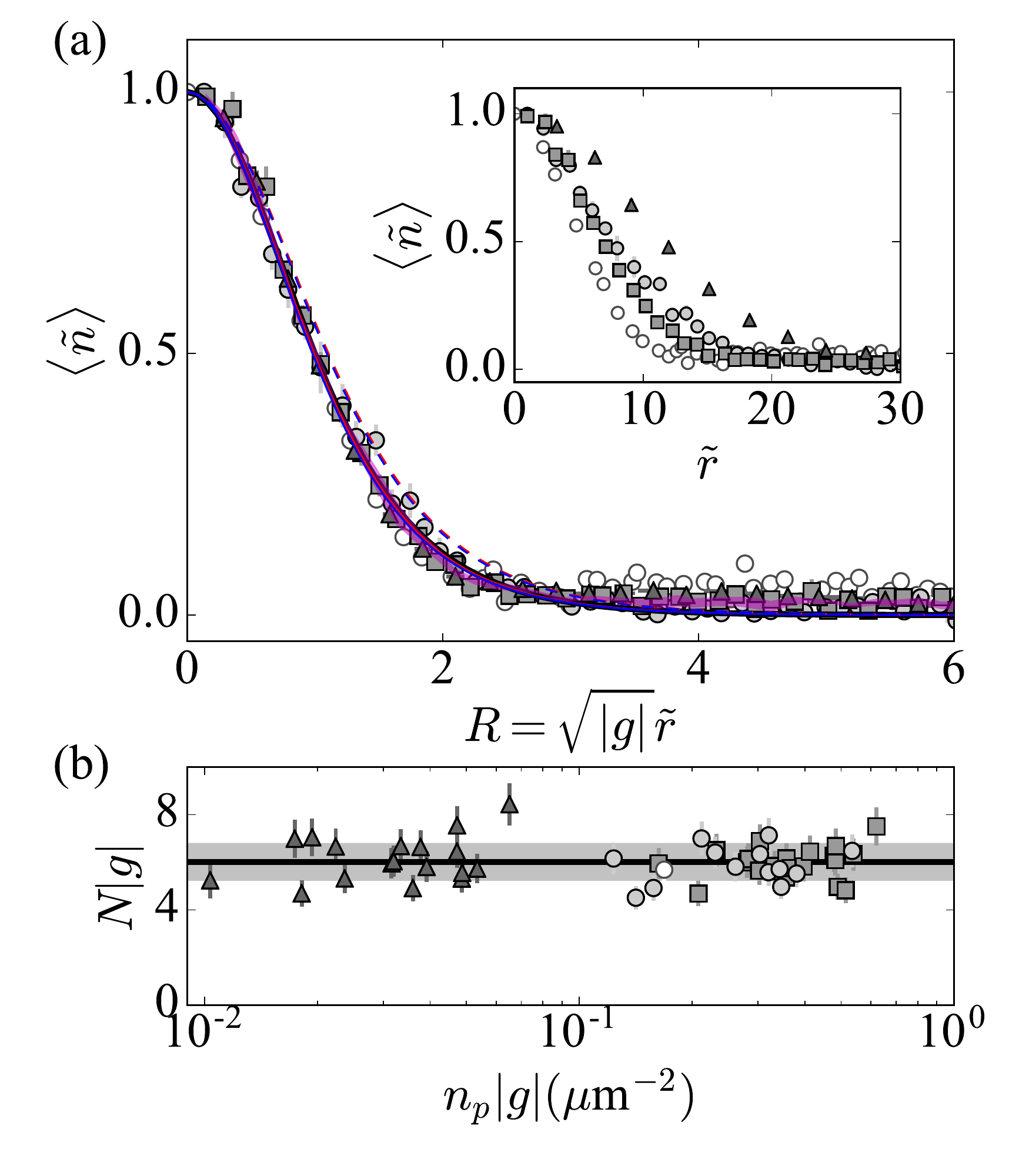}
\caption{Universal soliton density profile. (a) Filled symbols show different scale-invariant mean profiles $\mean{\tilde{n}}$ (inset), measured at interaction strengths $g \approx -0.0075$ (triangle), $-0.0170$ (circle), and $-0.0215$ (square), respectively. Open circles display a scaled density profile reported in Ref.~\cite{chen2020observation}, for $g\approx -0.034$ and with a fixed $n_p\approx 5/\mu\mathrm{m}^2$. These profiles collapse onto a single curve in the rescaled radial coordinate $R = \sqrt{|g|} \tilde{r}$, and the magenta band marks their mean with standard error. Collapsed solid curves are the universal Townes profile (black) and the solutions of full GPE with the MDDI term Eq.~(\ref{eq:mdi}), calculated using $g_c=-0.009$, $n_p = 1/\mu$m$^2$ (red) and $10/\mu$m$^2$ (blue), respectively, and rescaled using $g=g_c+2g_\mathrm{dd}$. For comparison, dashed curves show the same solutions rescaled using $g=g_c$. (b) Universal atom number $N |g| =\int \tn d\mathbf{R}$ using soliton profiles as in Fig.~\ref{fig:invariant} and integrated up to $R=4$. Solid line and gray band indicate the mean and standard deviation.}
\label{fig:universal}
\end{figure}

To form a single array of isolated 2D solitons, we reduce the initial width of a superfluid so that MI can only manifest along its long axis ($y$-axis). As shown in Fig.~\ref{fig:demo}(a), the sample has an initial peak density $n_i\approx 18/\mu$m$^2$, with a length $L\approx 65~\mu$m and a root-mean-square width $w\approx 3~\mu$m$~\lesssim \xi$, where $\xi=\pi/ \sqrt{2 n_i |g|}\approx 3.6~\mu$m is the half-wavelength of the most unstable mode in MI \cite{chen2020observation} when we quench to $g\approx -0.0215$. Following the interaction quench, arrays of isotropic solitary waves are observed to form near-deterministically in every sample [Fig.~\ref{fig:demo}(b)]. These well-separated solitary waves allow us to perform counting statistics (Fig.~\ref{fig:statistics}) and measure their density profiles. We confirm these solitary waves are Townes solitons by performing associated scaling tests (Figs.~\ref{fig:invariant} and \ref{fig:universal}). In another set of examples as shown in Fig.~\ref{fig:demo}(c-d), we prepare superfluids with much lower initial peak densities $n_i\approx 5/\mu$m$^2$, and quench the coupling constant to a less attractive value $g\approx -0.0075$. Arrays of solitons more than twice the size of those found in Fig.~\ref{fig:demo}(b) can be identified in (d).

In all examples shown in Fig.~\ref{fig:demo}, many solitons appear to be missing randomly from the observed arrays. This may be caused by imperfect soliton formation from MI, and the missing ones may have either dispersed or collapsed. In addition, collisions between neighboring solitons can trigger collapse and induce rapid loss \cite{nguyen2014collisions,chen2020observation}. In Fig.~\ref{fig:statistics}, we analyze soliton formation statistics from our quench recipe, using images as shown in Fig.~\ref{fig:demo}(b). In more than $98~\%$ of the samples analyzed, we find $N_s \geq 1$ total number of solitons [Fig.~\ref{fig:statistics}(a)]. Thanks to a nearly remnant-free background, we collect solitons of peak densities over a finite range from $n_p\sim 8/\mu$m$^{2}$ to $\sim 30/\mu$m$^{2}$ [Fig.~\ref{fig:statistics}(b)]. This allows us to study their density scaling behavior. On the other hand, the average peak density $\bar{n}_p \approx 20/\mu$m$^2$ [Fig.~\ref{fig:statistics}(c)] is comparable to the initial density $n_i\approx 18/\mu$m$^2$, and is approximately uniform along the sample. It is more likely to find solitons near the edge, as shown in the probability distribution $p(y)$ in Fig.~\ref{fig:statistics}(d), potentially due to a boundary effect that reduces soliton collision loss. We observe that low density samples as shown in Fig.~\ref{fig:demo}(d) generate solitons with peak density $2/\mu$m$^{2} \lesssim n_p$ $\lesssim 13/\mu$m$^{2}$.

We collect solitons of different sizes from our quenched samples to perform the scaling tests. In Fig.~\ref{fig:invariant}, we show sample soliton images, sorted with $n_p$ monotonically increasing from $7/\mu$m$^{2}$ to $30/\mu$m$^{2}$ for $g\approx -0.0215$ [in (a)] and from $1.5/\mu$m$^{2}$ to $9/\mu$m$^{2}$ for $g\approx -0.0075$ [in (b)]. The soliton size appears to monotonically decrease with respect to the increasing peak density, as shown in the radial density profiles $n(r)$ in Fig.~\ref{fig:invariant} insets.

We test the SI hypothesis by rescaling the density profiles $n(r)$ in a dimensionless form and search for a universal behavior. In Fig.~\ref{fig:invariant}, we plot the rescaled density $\tilde{n} = n/n_p$ as a function of the dimensionless radial position $\tilde{r}=\sqrt{n_p}r$. Indeed, despite a large variation in soliton size, we observe that all profiles measured at a fixed $g$ collapse onto a single curve. No significant deviation from the collapse behavior is observed at any $\tr$.

To quantify the goodness of the profile collapse and confirm SI, we evaluate the reduced chi-square $\chi_\nu^2 = \sum_i \left[\tilde{n}_i-\mean{\tilde{n}}_i\right]^2/\nu \sigma_i^2$ from $\sim 20$ rescaled profiles, where $\mean{\tilde{n}}$ is the mean profile, $\sigma_i$ is data uncertainty, and the index $i$ labels data points collected within a test radius, giving in total $\nu \approx 190$ degrees of freedom. At $g \approx -0.0215$ as in Fig.~\ref{fig:invariant}(a), we find $\chi_\nu^2 \approx 1.5$ for $\tr\lesssim 25$; for the profiles at $g\approx -0.0075$ as in Fig.~\ref{fig:invariant}(b), we obtain $\chi_\nu^2 \approx 1.4$ for $\tr\lesssim 35$. The chi-square test $\chi_\nu^2 \sim O(1)$ suggests a universal collapse and supports the SI hypothesis from these randomly collected solitons. Nevertheless, $\chi_\nu^2\gtrsim 1$ indicates that the standard deviation of collapsed profiles slightly exceeds the estimated measurement uncertainty. Since the statistical deviations from the mean profile show no clear dependence on soliton size or peak density [see also Fig.~\ref{fig:universal}(b)], the chi-square test suggests not all quench-induced solitary waves possess perfect scale-invariant profiles.

We now show that the scale-invariant density distributions measured at different attractive interactions can be further rescaled to display a universal waveform -- the Townes profile. Here, the coupling constant can be absorbed into the length scale factor $\lambda$ such that, when plotted in the rescaled coordinate $R = \sqrt{|g|} \tr$, the density displays a universal profile $\tilde{n} =|\phi(R)|^2$. The radial wave function $\phi(R)$ is the stationary solution of a dimensionless 2D Gross-Pitaevskii equation (GPE),
\begin{equation}
\tilde{H}\phi = - \frac{1}{2}\left(\frac{d^2\phi}{dR^2} + \frac{1}{R}\frac{d\phi}{d R}\right) - |\phi|^2\phi = \tilde{\mu} \phi\, , \label{GPE}
\end{equation}
where the scaled chemical potential $\tilde{\mu} = - 0.205$ is obtained while solving $\phi(R)$ \cite{SM}.

In Fig.~\ref{fig:universal}, we plot the measured scale-invariant mean density profiles $\mean{\tilde{n}}$ as a function of the rescaled radial position $R =\sqrt{|g|} \tr$. We find that four initially very different mean profiles (inset) measured at $|g| \approx (0.0075, 0.0170, 0.0215, 0.034)$, respectively, can collapse onto a universal curve in the rescaled coordinate, which agrees very well with the GPE solution $|\phi(R)|^2$; only a small deviation $\Delta\tilde{n}\lesssim 0.015$ becomes visible at $R\gtrsim 3$, where $|\phi(R)|^2 \lesssim 0.02$. This could result from a very low fraction of collision remnants in the horizontal plane or from barely overlapping tails of adjacent solitons, which has little influence on the universal scaling tests near the core region $R\lesssim 3$. Integrating the scaled density to $R=4$, we have estimated $\int \mean{\tn} d\mathbf{R}\approx 6.0\pm 0.8 \sim N_\mathrm{ts}|g|$, agreeing reasonably with theory [Fig.~\ref{fig:universal}(b)].

The observed universal scaling behavior is a remarkable manifestation of SI in 2D Bose gases effectively described by a mean field interaction Eq.~(\ref{GPE}). This universal behavior is also evidenced in Fig.~\ref{fig:universal} (b), where we plot the scaled atom number $N|g|$ of individual solitons as shown in Fig.~\ref{fig:invariant}. Almost all of them collapse to the universal number $N_\mathrm{ts}|g|$ to within the experiment uncertainty. The scaling behavior is tested with solitons of a nearly 60-fold difference in their peak interaction energies $\hbar \gamma = \hbar^2 n_p|g|/m$, where $\hbar$ is the reduced Planck constant, $m$ is the atomic mass, and $\gamma \approx 2\pi \times (0.85 - 49)~$Hz.

It is however worth noting that a non-negligible MDDI potential is present in our alkali cesium samples \cite{giovanazzi2002tuning,lahaye2009physics,olson2013effects}. Since a MDDI potential scales with the inter-atomic spacing as $1/r^{3}$, it could impact SI in a 2D Bose gas. For the effective 2D MDDI strength \cite{pedri2005two},
\begin{equation}
g_\mathrm{dd} = \frac{m}{\hbar^2}\frac{\mu_0\mu^2}{3\sqrt{2\pi} l_z}\, ,
\end{equation}
we find that $g_\mathrm{dd}\approx 0.00087$
is stronger than $-10\%$ of the smallest coupling constant $g\approx-0.0075$ explored, where $\mu_0$ is the vacuum permeability, $\mu \approx 0.75 \mu_B$ cesium magnetic dipole moment, and $\mu_B$ the Bohr magneton. It is thus necessary to examine the effect of MDDI in a GPE. The MDDI in our matter-wave solitons is in a highly oblate configuration, with spin polarized along the tightly confined $z$-axis. Integrating out wave function along this axis (assumed Gaussian), the rescaled MDDI Hamiltonian can be conveniently expressed as the following inverse Fourier transform \cite{pedri2005two,fischer2006stability, mishra2016dipolar}:
\begin{equation}
\tilde{H}_\mathrm{dd} = \frac{g_\mathrm{dd}}{|g_c|}\int \frac{d\mathbf{k}}{(2\pi)^2}e^{ik R \cos\theta_k}h_\mathrm{dd}\left(\sqrt{\frac{n_p |g_c| }{2}}kl_z\right) \tilde{n}(\mathbf{k})\, ,\label{eq:mdi}
\end{equation}
where we define $g_\mathrm{c}$ as the bare contact coupling constant, $\tilde{n}(\mathbf{k})$ is the Fourier transform of the rescaled density profile $\tilde{n}(\mathbf{R})=|\phi(R)|^2$, and $h_\mathrm{dd}$ is the MDDI function that can potentially break SI \cite{SM}. However, in the limit $\sqrt{n_p |g|}l_z \ll 1$, $h_\mathrm{dd} \approx 2$ is approximately constant within a finite $k$-range until $\tilde{n}(\mathbf{k})$ vanishes. Equation~(\ref{eq:mdi}) thus transforms back to an effective contact interaction Hamiltonian:
\begin{equation}
\tilde{H}_\mathrm{dd} \approx 2 \frac{g_\mathrm{dd}}{|g_c|} |\phi(R)|^2 \, .
\end{equation}
This argument generally applies to weakly interacting 2D gases whose lateral size $w \gg l_z$ \cite{SM,mishra2016dipolar}. As such, the full Hamiltonian in a modified GPE, $\tilde{H}+\tilde{H}_\mathrm{dd}$, can be effectively recast into $\tilde{H}$ in Eq.~(\ref{GPE}) by rescaling the coordinate $R$ using $g = g_\mathrm{c} + 2 g_\mathrm{dd}$.

We numerically confirm SI with the MDDI in our quasi-2D samples that have a small but finite $l_z\approx 184~$nm, giving $0.02 \lesssim \sqrt{n_g|g|}l_z \lesssim 0.15$ \footnote{More precisely, one should call this quasi-SI in quasi-2D samples with finite $l_z$, as there exists small differences in the rescaled profiles well below typical experiment uncertainty.}. As shown in Fig.~\ref{fig:universal}, sample numerical solutions at $g_c=-0.009$ collapse well to the universal Townes profile if we rescale the radial coordinate $R$ using $g = g_\mathrm{c} + 2 g_\mathrm{dd} \approx -0.0073$, which includes the MDDI shift.

The good agreement between our measurement results and the properly rescaled numerical solutions suggests our coupling constant $g$, which is evaluated using a calibrated scattering length, is already shifted by the MDDI \cite{SM,pollack2009extreme,olson2013effects}. This is likely the case, as our calibration procedure performed in a quasi-2D trap cannot discern the effect of MDDI from that of a two-body contact interaction \cite{SM}. We conclude that the scaling tests performed in Figs.~\ref{fig:invariant} and \ref{fig:universal} confirm SI with the inclusion of a weak MDDI contribution in our quasi-2D geometry.

In summary, we demonstrate a near-deterministic method to form 2D matter-wave solitons and test the scaling symmetry in attractive 2D Bose gases previously inaccessible to other experiments. We show that SI manifests robustly through an unstable many-body state, formed remarkably from out-of-equilibrium quench dynamics \cite{chen2020observation}. In particular, our observation confirms that the Townes profile not only manifests in a self-similar nonlinear wave collapse, as partially observed in Ref.~\cite{moll2003self}, it is also a prevalent SI profile in solitary waves formed from a modulational instability. The observed universal scaling behavior is under the influence of a non-negligible MDDI potential, which nevertheless imposes no influence on SI in a quasi-2D geometry. A recent study also reveals the insensitivity in the size and shape of a 2D superfluid to the MDDI \cite{zou2020magnetic}. Our recipe for instability-induced soliton formation may be further explored in a SI-breaking scenario, for example, through crossover to an MDDI-dominating regime \cite{pedri2005two,lahaye2009physics}, either by tuning to a much smaller contact coupling $g_c$ \cite{pollack2009extreme} or with a dipolar quantum gas \cite{griesmaier2005bose,lu2011strongly,aikawa2012bose,petter2019probing}. Furthermore, our scaling analysis may be extended to test the dynamics of stronger attractive 2D Bose gases, where quantum correlations may begin to play an important role, such as those discussed in quantum droplets \cite{petrov2015quantum,ferrier2016observation,chomaz2016quantum,semeghini2018self,cabrera2018quantum,cheiney2018bright}.

\begin{acknowledgments}
This work is supported by the NSF (Grant \# PHY-1848316), the W. M. Keck Foundation, and the DOE QuantISED program (Grant \# DE-SC0019202).
\end{acknowledgments}

\onecolumngrid
\appendix*
\renewcommand{\thesubsection}{S\arabic{subsection}}
\renewcommand{\figurename}{Fig.}
\renewcommand{\thefigure}{SM\arabic{figure}}
\setcounter{figure}{0}
\renewcommand{\theequation}{S\arabic{equation}}
\setcounter{equation}{0}
\setcounter{page}{1}

\section*{Supplemental Material}
\subsection{Magnetic two-body interaction tuning} \label{secSM:tuning}
We tune the cesium scattering length by applying a uniform bias magnetic field perpendicular to the $x$-$y$ plane to access a magnetic Feshbach resonance \cite{chin2010feshbach,kraemer2006evidence}. We identify zero scattering length at the magnetic field $B=17.120(6)$G, by minimizing superfluid in-situ size as well as the expansion rate in a 2D time-of-flight. We then adopt the formula \cite{kraemer2006evidence} $a(B) = (1722 + 1.52B/G)\left(1 - \frac{\Delta B}{B/G - B_0}\right)$ for the scattering length conversion, where $\Delta B = 28.72$ and $B_0=-11.60$ is adjusted to shift the zero-crossing to the measured value. The interaction strength is determined as $g =\sqrt{8\pi}a/l_z$, where $l_z = 184~$nm is the vertical harmonic oscillator length. The uncertainty ($\pm \delta g$) in $g$ is primarily contributed by the uncertainty in the magnetic field at the scattering length zero-crossing. Within the range of our reported negative interaction strengths $-0.0075 \geq g \geq -0.022$, we have $\delta g \approx 0.0005$. 

While we have calibrated the coupling constant $g$ with relatively small uncertainty near zero-crossing, our measurement method does not distinguish a small offset contribution from the magnetic dipole-dipole interaction (MDDI). The 2D superfluid samples adopted in this calibration have in-situ widths of $w > 7~\mu$m much larger than the size $l_z =184~$nm along the tightly confining axis. This large aspect ratio and our magnetic field orientation (perpendicular to the 2D plane) makes the mean field effect of the MDDI effectively a contact-like interaction, contributing to a shift in the calibrated coupling constant 
\begin{equation}
g \approx g_c + 2g_\mathrm{dd}
\end{equation}
where $g_c$ is the bare coupling constant of the contact interaction and $g_\mathrm{dd} \approx 0.00087$ is the MDDI coupling constant of atomic cesium confined in the quasi-2D trap; see discussions below.

\subsection{Scale-invariant 2D solitons} 
In the following sections, we evaluate the stationary 2D matter-wave density profile and consider the presence of a MDDI. We start by considering the Gross-Pitaevskii equation (GPE) with a coupling constant $g_c<0$ for the contact interaction potential. Due to strong vertical confinement along the $z$-axis, the vibrational level spacing $\hbar \omega_z \gg \frac{\hbar^2}{m}n_p |g_c|$ is much larger than the absolute value of the interaction energy, where $n_p$ is the peak density, $\hbar$ the reduced Planck constant, and $m$ the atomic mass. The atomic wave function is frozen to the harmonic ground state along the $z$-axis. Integrating out the $z$-dependence in the GPE and assuming the wave function is isotropic in the $x$-$y$ plane, we have
\begin{equation}
H \psi =- \frac{\hbar^2}{2 m }\left(\frac{d^2\psi}{dr^2} + \frac{1}{r}\frac{d\psi}{d r}\right) + \frac{\hbar^2g_c}{m} |\psi|^2\psi = \mu \psi\, , \label{eqSM:GPE}
\end{equation}
where $\psi(\mathbf{r})=\psi(r)$ is the wave function that only has a radial dependence and $n(r)=|\psi(r)|^2$ is the radial density profile. We rescale Eq.~(\ref{eqSM:GPE}) using 
\begin{align}
&\mathbf{R}=\sqrt{|g_c|n(0)}\mathbf{r} \, , \label{eqSM:rescaler}\\
&\frac{\psi(\mathbf{r})}{\sqrt{n(0)}} \rightarrow \phi(\mathbf{R})\, , \label{eqSM:rescale}
\end{align}
and arrive at a scale-invariant GPE
\begin{equation}
\tilde{H} \phi =- \frac{1}{2 }\left(\frac{d^2\phi}{d R^2} + \frac{1}{R}\frac{d\phi}{d R}\right) - |\phi|^2\phi = \tilde{\mu} \phi\, . \label{eqSM:scaledGPE}
\end{equation}
The above equation can be numerically solved. The solution gives a chemical potential 
\begin{equation}
\tilde{\mu}=\tilde{\mu}_\mathrm{ts} = -0.205\, . \label{eqSM:muts}
\end{equation}
We call the resulting scale-invariant solution, $|\phi_\mathrm{ts}(R)|^2$, the Townes profile. Assigning a peak density $n_p=n(0)$ and a coupling constant $g_c$, a Townes soliton must have the density profile 
\begin{equation}
n(r) = n_p \left|\phi_\mathrm{ts}\left(\sqrt{n_p|g_c|}r\right)\right|^2\, .
\end{equation}

\subsection{Effect of the MDDI on 2D scale invariance}
We now consider the impact on scale invariance with the addition of an MDDI term in the GPE
\begin{equation}
H_\mathrm{dd} = \int d\mathbf{r}'V_\mathrm{dd}(\mathbf{r}-\mathbf{r}')|\psi(\mathbf{r}')|^2\, ,
\end{equation}
where $V_\mathrm{dd}(r,\theta) = \frac{\mu_0 \mu^2}{4\pi} (1-3\cos^2\theta)/r^3$ is the magnetic dipole-dipole potential, $\mu_0$ is the vacuum permeability, $\mu\approx 0.75\mu_B$ is the magnetic moment of cesium near scattering length zero-crossing, and $\mu_B$ is the Bohr magneton. The above convolution integral can be expressed in the Fourier space, where the $k_z$ dependence can be integrated out. We have
\begin{equation}
H_\mathrm{dd}= \frac{\hbar^2g_\mathrm{dd}}{m}\int \frac{d\mathbf{k}}{(2\pi)^2}e^{ikr\cos\theta_k}h_\mathrm{dd}\left(\frac{k l_z}{\sqrt{2}}\right) n(\mathbf{k})\, , \label{eqSM:hdd}
\end{equation}
where 
\begin{equation}
g_\mathrm{dd} = \frac{m}{\hbar^2}\frac{\mu_0\mu^2}{3\sqrt{2\pi} l_z}
\end{equation}
is the 2D MDDI coupling strength and $n(\mathbf{k})$ is the 2D Fourier transform of the density profile $|\psi(\mathbf{r})|^2$. The MDDI function reads \cite{pedri2005two,mishra2016dipolar}
\begin{equation}
h_\mathrm{dd}(x) = (3\cos^2\alpha -1) + 3 \sqrt{\pi} x e^{x^2} \mathrm{efrc}\left(x\right) \left(\sin^2\alpha \cos^2\theta_k -\cos^2\alpha\right)\, ,
\end{equation}
where $\mathrm{efrc}(x)$ is the complementary error function and $\alpha$ is the angle between the spin axis and the tight-confining $z$-axis. In our experimental setup, $\alpha = 0$ and $h_\mathrm{dd}$ simplifies to 
\begin{equation}
h_\mathrm{dd}\left(\frac{k l_z}{\sqrt{2}}\right)= 2 - 3 \sqrt{\pi}\left(\frac{k l_z}{\sqrt{2}}\right)e^{k^2l_z^2/2}\mathrm{efrc}\left(\frac{k l_z}{\sqrt{2}}\right)\, .
\end{equation}

We now express the full 2D Hamiltonian in the rescaled unit according to Eq.~(\ref{eqSM:rescale})
\begin{equation}
\left(\tilde{H} + \tilde{H}_\mathrm{dd}\right) \phi - \frac{1}{2 }\left(\frac{d^2\phi}{d R^2} + \frac{1}{R}\frac{d\phi}{d R}\right) - |\phi|^2\phi + \frac{g_\mathrm{dd}}{|g_c|} \int \frac{d\mathbf{k}}{(2\pi)^2}e^{i k R \cos\theta_k}h_\mathrm{dd}\left(\sqrt{\frac{n_p|g_c|}{2}}k l_z\right) \tilde{n}(\mathbf{k})\phi \tilde{\mu}\phi\, ,\label{eqSM:MDDIH}
\end{equation}
where $\tilde{n}(\mathbf{k})$ is the dimensionless 2D Fourier transform of the rescaled density profile $|\phi(\mathbf{R})|^2$. 

\subsubsection*{Scale invariance in deep 2D limit}
We first consider the deep 2D limit with small $ l_z \ll w$, where $w$ is the characteristic horizontal size of the sample, and $\sqrt{n_p|g_c|}l_z \ll 1$. In this case, $\tilde{n}(\mathbf{k})$ is non-vanishing only when $k \lesssim O(2\pi/w\sqrt{n_p|g_c|})$, where $h_\mathrm{dd}(\sqrt{\frac{n_p|g_c|}{2}}k l_z) = 2$ remains a constant. The MDDI Hamiltoanian in Eq.~(\ref{eqSM:MDDIH}) thus gives
\begin{equation}
\tilde{H}_\mathrm{dd} = 2\frac{g_\mathrm{dd}}{|g_c|}|\phi|^2\, ,
\end{equation}
which carries the same form of a contact interaction term. Equation~(\ref{eqSM:MDDIH}) can thus be recast into the exact same form of Eq.~(\ref{eqSM:scaledGPE}) by rescaling using
\begin{equation}
\mathbf{R} =\sqrt{n_p |g|}\mathbf{r}\, ,\label{eqSM:rddi}
\end{equation}
where the bare coupling constant in Eq.~(\ref{eqSM:rescaler}) is replaces by 
\begin{equation}
g = g_c + 2g_\mathrm{dd}\, . \label{eqSM:shift}
\end{equation}
The stationary solution of a 2D matter-wave with $g<0$ remains to be that of a scale-invariant Townes profile. The solution has a chemical potential $\tilde{\mu}$ that relates to the solution $\tilde{\mu}_\mathrm{ts}$ of Eq.~(\ref{eqSM:scaledGPE}) as
\begin{equation}
\tilde{\mu} =\frac{g}{g_c}\tilde{\mu}_\mathrm{ts}\, . \label{eqSM:mod_mu}
\end{equation}

\begin{figure}[t]
\includegraphics[width=1\columnwidth]{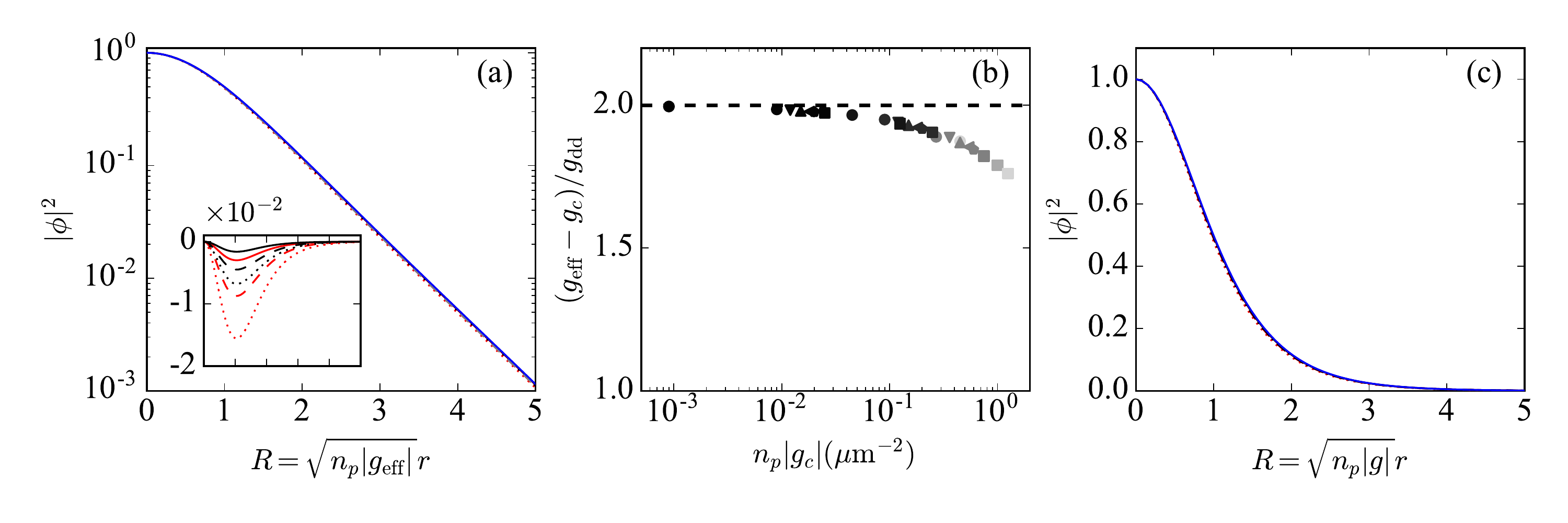}
\caption{Quasi-scale invariance in quasi-2D matter-wave solitons with the MDDI. (a) Scaled density profiles $|\phi(R)|^2$ evaluated using the full Hamiltonian Eq.~(\ref{eqSM:MDDIH}) with $l_z=184~$nm, bare coupling constants $g_c=-0.009$ (red curves), $-0.025$ (black curves), and with peak densities $n_p=1/\mu\rm{m}^2$ (solid), $10/\mu\rm{m}^2$ (dashed), and $30/\mu\rm{m}^2$ (dotted), respectively. Blue curve shows the Townes profile $|\phi_\mathrm{ts} (R)|^2$ without the MDDI. Inset plots the difference $|\phi(R)|^2-|\phi_\mathrm{ts}(R)|^2$. (b) Filled symbols show the deviation of $g_\mathrm{eff}$ from the bare coupling constant $g_c$, normalized by $g_\mathrm{dd}$. Dashed line marks the 2D limit ($g-g_c=2g_\mathrm{dd}$); $g_\mathrm{eff}$ is evaluated using $g_c = -0.009$ (circles), $-0.012$ (down triangles), $-0.015$ (up triangles), $-0.018$ (left triangles), $-0.02$ (pentagons), and $-0.025$ (squares), respectively, and at various peak densities $n_p = 0.1/\mu$m$^2 \sim 50/\mu$m$^2$ (dark to light gray). (c) Density profiles $|\phi(R)|^2$ as in (a), but with the radial coordinate rescaled using $g=g_c+2g_\mathrm{dd}$ and plotted in linear scale.}
\label{figSM:scale}
\end{figure}

\subsubsection*{Quasi-scale invariance}
In the present experiment, we have $0.02 \lesssim \sqrt{n_g|g|}l_z \lesssim 0.15$ and $w \gtrsim 2~\mu$m$> l_z$ approximating the 2D limit. Here, we numerically confirm an effective scale-invariant scaling behavior (quasi-scale invariance) for stationary states realized in our experiment. To see if the scaling behavior is effectively preserved, we numerically solve for the soliton density profiles by finding the solutions to the integro-differential Eq.~(\ref{eqSM:MDDIH}). Firstly, we obtain the chemical potential $\tilde{\mu}$ and define the effective 2D coupling constant following the relation Eq.~(\ref{eqSM:mod_mu}):
\begin{equation}
g_\mathrm{eff} = \frac{\tilde{\mu}}{\tilde{\mu}_\mathrm{ts}} g_c\, .
\end{equation}
which should approach $g=g_c+2g_\mathrm{dd}$ in the 2D limit. We then rescale the radial coordinate of the stationary density profile according to Eq.~(\ref{eqSM:rddi}) using the effective coupling constant $g_\mathrm{eff}$. Figure~\ref{figSM:scale}(a) plots the rescaled profiles $|\phi(R)|^2$ evaluated at various $(n_p,g_c)$ around and beyond our experiment parameters. Indeed, the stationary profiles collapse very well to the universal Townes profile, with much less than $<2\%$ deviation (relative to the peak density) over the entire density profile. In Fig.~\ref{figSM:scale}(b), we calculate the shift in $g_\mathrm{eff}$ relative to the bare coupling constant $g_c$. The shift approaches the 2D limit ($2g_\mathrm{dd}$) quite well, even after we increase the interaction parameter $n_p|g_c|$ by three orders of magnitude from $n_p|g_c|= 10^{-3}/\mu$m$^2$ up to $n_p|g_c|= 1/\mu$m$^2$, where only a small deviation $\Delta g \approx 0.25g_\mathrm{dd} \approx 0.00022 \ll g_\mathrm{eff}$ occurs. In Fig.~\ref{figSM:scale}(c), we plot the same density profiles $|\phi(R)|^2$ but with the radial coordinate $R$ rescaled using $g=g_c+2g_\mathrm{dd}$, as this should be closer to the scaling performed in our experiment (see Fig.~\ref{fig:universal} and discussion in Sec.~\ref{secSM:tuning}). The Townes profile remains to be an excellent universal description for the rescaled density profiles.

\newpage
\end{document}